**GENERAL**

**Regular**

# OC-PM: analyzing object-centric event logs and process models

Alessandro Berti[1] · Wil M. P. van der Aalst[1]

**Abstract**
Object-centric process mining is a novel branch of process mining that aims to analyze event data from mainstream information systems (such as SAP) more naturally, without being forced to form mutually exclusive groups of events with the specification of a case notion. The development of object-centric process mining is related to exploiting object-centric event logs, which includes exploring and filtering the behavior contained in the logs and constructing process models which can encode the behavior of different classes of objects and their interactions (which can be discovered from object-centric event logs). This paper aims to provide a broad look at the exploration and processing of object-centric event logs to discover information related to the lifecycle of the different objects composing the event log. Also, comprehensive tool support (OC-PM) implementing the proposed techniques is described in the paper.

**Keywords** Object-centric process mining · Object-centric event logs · Object-centric process discovery · Object-centric conformance checking

## 1 Introduction

Process mining [1] is a branch of data science providing techniques to exploit the event data that support the execution of business processes. Different areas exist in process mining, such as process discovery (the discovery of a process model from an event log), conformance checking (comparing the behavior described in the event log with a process model), process enhancement (enriching models with information related to time and data), machine learning (such as root cause analysis and the prediction of the next activity/remaining time).

Mainstream process mining techniques start from an *event log*, i.e., a collection of events extracted from the databases supporting the process execution. In such event logs, a *case* is a collection of events related to a particular process execution. For example, in a sales order management system, a case may refer to all the events related to the creation and confirmation of the order, collecting and packing the different order items, the delivery, and invoicing. For such a system, establishing a case notion can lead to the known *convergence* and *divergence* problems [2]. We have a convergence problem when the same event is related to different cases. In event log formats such as XES[1], this leads to replicating the same event. We have a divergence problem when a case contains different instances of the same activity. For example, a sales order contains different instances of the collection and the packing of the order items. Mainstream process mining techniques (e.g., process discovery and conformance checking) use the order of the events inside the log cases. However, the quality of the output is affected by the convergence and divergence problems.

*Object-centric event logs* have been proposed to resolve the convergence and divergence problems. Object-centric event logs are a novel representation of the event data in the information systems, where each event is related to different objects of different types. An informal representation of an object-centric event log is contained in Table 1. The event log contains eight events. The first event, having activity "create order", is related to an object of type order ($o_1$) and two objects of type item ($i_1$ and $i_2$). Moreover, some attributes at

✉ Alessandro Berti
a.berti@pads.rwth-aachen.de

Wil M. P. van der Aalst
wvdaalst@pads.rwth-aachen.de

[1] Process and Data Science Group, RWTH Aachen University, Ahornstrasse 55, 52074 Aachen, NRW, Germany

---

[1] https://xes-standard.org/

Springer

**Table 1** Informal representation of the events of an object-centric event log. Each row (except the header) represents an event

| Id | Activity | Timestamp | Item | Order | Package | Delivery | Resource | Prepaid-amount | weight | Total-weight |
|---|---|---|---|---|---|---|---|---|---|---|
| $e_1$ | Place order | 2020-07-09 08:20:01.527+01:00 | { $i_1, i_2$ } | { $o_1$ } | | | Alessandro | 200.0 | | |
| $e_2$ | Check availability | 2020-07-09 08:21:01.527+01:00 | { $i_1$ } | | | | Anahita | | 10.0 | |
| $e_3$ | Check availability | 2020-07-09 08:22:01.527+01:00 | { $i_2$ } | | | | Anahita | | 20.0 | |
| $e_4$ | Create package | 2020-07-09 08:21:01.527+01:00 | { $i_1$ } | | { $p_1$ } | | Miriam | | 10.0 | |
| $e_5$ | Create package | 2020-07-09 08:21:01.527+01:00 | { $i_2$ } | | { $p_2$ } | | Tobias | | 20.0 | |
| $e_6$ | Load package | 2020-07-09 08:23:01.527+01:00 | | | { $p_1, p_2$ } | { $d_1$ } | Marco | | | 30.0 |
| $e_7$ | Delivery successful | 2020-07-09 08:23:01.527+01:00 | | | | { $d_1$ } | Marco | | | 30.0 |
| $e_8$ | Order completed | 2020-07-09 08:24:01.527+01:00 | | { $o_1$ } | | | Alessandro | | | |

the event level are described (for example, *prepaid-amount* having a value 200.0 for the first event). Object-centric event logs describe the lifecycle of different objects without leading to the convergence problem since an event can be related to different objects. Also, the divergence problem is avoided since we avoid specifying a single case notion.

In an object-centric event log, it is also possible to specify the values for some attributes of the objects. This is represented in Table 2: object $o_1$, having type order, is sold to the customer Apple at the cost of 3500. To exploit the information contained in object-centric event logs, new process mining techniques are required. This leads to the development of *object-centric process mining* techniques. These exploit the lifecycle of the objects and the relationships between the objects to provide insights on the execution of a business process and check the actual execution logs.

However, the discipline is still in an early development stage. While some approaches have been proposed for process discovery and conformance checking in an object-centric setting (see Sect. 5), some important aspects, such as the exploration (being able to visualize the event log and focus on some dimensions) and filtering (restrict the behavior of the log to a subset of events/objects) of object-centric event logs have been marginally explored. Moreover, a comprehensive description of the annotations for the elements of object-centric process models (such as the frequency of nodes and edges) is missing. The paper aims to propose a set of object-centric process mining techniques to bridge the gap between traditional and object-centric process mining, including:

- The exploration of object-centric event logs: its events and the *lifecycle* of its objects.
- The filtering possibilities on object-centric event logs.
- Provide the automatic discovery of object-centric process models with different complexity (less/more activities, less/more edges).
- Describe meaningful annotations (number of events/ objects) at the activity and the edge level.
- Provide some conformance metrics on object-centric event logs.

Also, the paper describes the OC-PM tool(s) for object-centric process mining, providing the proposed techniques as a web-based interface and as a plugin of the ProM framework. These are available at the address https://ocpm.info. Tool support in object-centric process mining is currently limited to a set of library and ad-hoc tools (for example, [3] for the discovery of variants), hence the importance of comprehensive tool support covering a good part of the lifecycle of an object-centric process mining analysis.

The rest of the paper is organized as follows. Section 2 describes the information on event logs and directly-follows graphs needed to understand the paper. Section 3 presents some operations on object-centric event logs (flattening, lifecycle, filtering), the discovery of object-centric directly-follows multigraphs, and several annotations at the activity and the edge level. Section 4 presents the tools supporting the paper. Section 5 presents the related work on object-centric process mining. Finally, Sect. 6 concludes the paper.

**Table 2** Informal representation of the objects of an object-centric event log. Each row (except the header) represents an object

| Id | Type | Customer | Costs | Color | Size | Ensured | Priority |
|---|---|---|---|---|---|---|---|
| $o_1$ | Order | Apple | 3500.0 | | | | |
| $i_1$ | Item | | | Orange | Big | | |
| $i_2$ | Item | | | Green | Small | | |
| $p_1$ | Package | | | | | Yes | |
| $p_2$ | Package | | | | | No | |
| $d_1$ | Delivery | | | | | | High |

## 2 Background

This section introduces the basic knowledge (event logs, directly-follows graph) needed to understand the rest of the paper.

### 2.1 Traditional event log

"Traditional" event logs, used by mainstream process mining techniques, are a collection of events and cases. A case is a collection of events of the same process execution. For example, in a ticket management system, a case contains events for the creation, the resolution, and the closure of the ticket. To introduce a definition of traditional event logs, we introduce some universes (event identifiers, case identifiers, activities) in Def. 1 and then the definition of traditional event log in Def. 2.

**Definition 1** (Generic Universes) Below are the universes used in the definition of traditional (and object-centric) event logs:

- $U_e$ is the universe of event identifiers. *Example:* $U_e = \{e_1, e_2, e_3, \ldots\}$
- $U_c$ is the universe of case identifiers. *Example:* $U_c = \{c_1, c_2, \ldots\}$
- $U_{act}$ is the universe of activities. *Example:* $U_{act} = \{$*place order*, *check availability*, $\ldots\}$
- $U_{timest}$ is the universe of timestamps. *Example:* $U_{timest} = \{$ *2020-07-09T08:21:01.527+01:00*, $\ldots\}$ We assume $U_{timest}$ to be totally ordered. Moreover, a difference $-$ operation is defined for timestamps as the number of seconds separating the subtrahend from the minuend.

**Definition 2** (Traditional Event Log) A *traditional* event log is a tuple $TL = (E, \pi_{act}, \pi_{time}, \pi_{case}, \leq)$ where:

- $E \subseteq U_e$ is a set of events.
- $\pi_{act} : E \to U_{act}$ associates an activity to each event.
- $\pi_{time} : E \to U_{timest}$ associates a timestamp to each event.
- $\pi_{case} : E \to \mathcal{P}(U_c) \setminus \{\emptyset\}$ associates a non-empty set of cases to each event.
- $\leq \; \subseteq E \times E$ is a total order on $E$.

An important consideration is that, in Def. 2, each event can be associated with several cases. In traditional event log formats such as XES[2], cases are primary-level objects, so if an event belongs to different cases, it is going to be replicated in the serialization.

The operations introduced in Def. 3 can be defined on an event log.

**Definition 3** (Operations on an Event Log) Given a traditional event log $TL = (E, \pi_{act}, \pi_{time}, \pi_{case}, \leq)$, we define the following operations:

- $\pi_{act}(TL) = \{\pi_{act}(e) \mid e \in E\}$
- $\pi_{case}(TL) = \cup_{e \in E} \; \pi_{case}(e)$
- For $c \in \pi_{case}(TL)$, $case_{TL}(c) = \langle e_1, \ldots, e_n \rangle$ where:
  - $\{e_1, \ldots, e_n\} = \{e \in E \mid c \in \pi_{case}(e)\}$
  - $\forall_{1 \leq i < n} \; e_i < e_{i+1}$
- Given $c \in \pi_{case}(TL)$ and $case_{TL}(c) = \langle e_1, \ldots, e_n \rangle$:
  - $trace_{TL}(c) = \langle \pi_{act}(e_1), \ldots, \pi_{act}(e_n) \rangle$
  - $start_{TL}(c) = \pi_{act}(e_1)$.
  - $end_{TL}(c) = \pi_{act}(e_n)$.
- $\pi_{start}(TL) = \{start_{TL}(c) \mid c \in \pi_{case}(TL)\}$ is the set of start activities.
- $\pi_{end}(TL) = \{end_{TL}(c) \mid c \in \pi_{case}(TL)\}$ is the set of end activities.

---
[2] http://www.processmining.org/logs/xes

## 2.2 Directly-follows graphs

A *directly-follows graph* (DFG) is a simple process model describing the directly-follows relationship between the different activities of a process. In Def. 4, we introduce a formal definition of DFG. The definition comes with a frequency measure on the nodes and a frequency measure on the edges of the DFG. This identifies the most and the least used paths in the process model.

**Definition 4** (Directly-Follows Graph) A *directly-follows graph* is a tuple $(A, F, \pi_{freqn}, \pi_{freqe})$ where:

- $A \subseteq U_{act}$ is a set of activities.
- $\triangleright$ is the start node of the graph, $\square$ is the end node of the graph.
- $F \subseteq (\{\triangleright\} \cup A) \times (A \cup \{\square\})$ is the set of edges.
- $\pi_{freqn} : A \nrightarrow \mathbb{N}$ is a frequency measure on the nodes.
- $\pi_{freqe} : F \nrightarrow \mathbb{N}$ is a frequency measure on the edges.

In the previous definition, we use $\nrightarrow$ as a symbol telling that a subset of elements of the domain (*A* and *F* respectively) is mapped to an element of the image ($\mathbb{N}$). We can discover a directly-follows graph from a traditional event log. This is introduced in Def. 5.

**Definition 5** (Discovery of a Directly-Follows Graph) Let $TL = (E, \pi_{act}, \pi_{time}, \pi_{case}, \leq)$ be a traditional event log. We define the following discovery operation:

$$DFG(TL) = (A, F, \pi_{freqn}, \pi_{freqe})$$

where:

- $A = \pi_{act}(TL)$ and

$$F = \{(\triangleright, \text{start}_{TL}(c)), (\text{end}_{TL}(c), \square) \mid c \in \pi_{case}(TL)\} \cup$$
$$\cup_{c \in \pi_{case}(TL), \text{trace}_{TL}(c) = \langle a_1, \ldots, a_n \rangle} \{(a_i, a_{i+1}) \mid 1 \leq i < n\}$$

- For $a \in A$, $\pi_{freqn}(a) = |\{e \in E \mid \pi_{act}(e) = a\}|$
- For $(\triangleright, a) \in F, \pi_{freqe}(\triangleright, a) = |\{c \in \pi_{case}(TL) \mid \text{start}_{TL}(c) = a\}|$
- For $(a, \square) \in F, \pi_{freqe}(a, \square) = |\{c \in \pi_{case}(TL) \mid \text{end}_{TL}(c) = a\}|$
- For $(a, b) \in F \cap (A \times A)$,

$$\pi_{freqe}(a, b) = \sum_{\substack{c \in \pi_{case}(TL), \\ \text{trace}_{TL}(c) = \langle a_1, \ldots, a_n \rangle}}$$
$$|\{i \in \mathbb{N} \mid 1 \leq i < n \land a_i = a \land a_{i+1} = b\}|$$

Given Def. 5, we can define two operators: the *projection on the set of activities* ($\pi_A((A, F, \pi_{freqn}, \pi_{freqe})) = A$) and the *set of edges* ($\pi_F((A, F, \pi_{freqn}, \pi_{freqe})) = F$).

The directly-follows graphs are the building blocks for some object-centric process models introduced in Sect. 3.3. As seen in Def. 5, they can be straightforwardly discovered from event logs, and they can be easily filtered based on a threshold on the frequency of activities and edges.

## 3 Approach

The approach section is composed of different subsections, analyzing different techniques to exploit object-centric event logs. We start with the definition of object-centric event logs and the proposition of the OCEL format for the storage of object-centric event logs. The flattening operation (which projects the object-centric event log to a traditional event log with the choice of a case notion) is introduced, as it is essential for process discovery purposes. Then, some filtering operations (activities, paths, endpoints, timeframe, object types) are proposed on top of object-centric event logs.

In Sect. 3.3, an object-centric process model (the object-centric directly-follows multigraph) is defined that can be discovered straightforwardly from an object-centric event log using the aforementioned flattening operation. Then, some generic metrics on object-centric event logs are introduced, which can be used to annotate the object-centric process models.

Finally, some model-independent conformance checking techniques are introduced, which can be applied to object-centric event logs.

### 3.1 Object-centric event log and flattening

The starting point of an object-centric process mining analysis lies in an object-centric event log. In object-centric event logs, we assume that each event is related to different objects of different types. Moreover, some other attributes are associated with the events and the objects of the log. Def. 6 introduces some universes that are necessary for the formal definition of object-centric event log. The definition has also been introduced in [4].

**Definition 6** (Universes (for OCEL)) Below are the universes used in the formal definition of object-centric event logs:

- $U_{att}$ is the universe of attribute names. *Example:* $U_{att} = \{\textit{resource, weight, \ldots}\}$
- $U_{val}$ is the universe of attribute values. *Example:* $U_{val} = \{\textit{500, 1000, Mike, \ldots}\}$
- $U_{typ}$ is the universe of attribute types. *Example:* $U_{typ} = \{\textit{string, integer, float, \ldots}\}$
- $U_o$ is the universe of object identifiers. *Example:* $U_o = \{o_1, i_1, \ldots\}$

- $U_{ot}$ is the universe of objects types. *Example:* $U_{ot} = \{$ *order*, *item*, ...$\}$

Def 7 introduces the formal definition of object-centric event log.

**Definition 7** (Object-Centric Event Log) An object-centric event log is a tuple
$L = (E, AN, AV, AT, OT, O, \pi_{typ}, \pi_{act}, \pi_{time}, \pi_{vmap}, \pi_{omap}, \pi_{otyp}, \pi_{ovmap}, \leq)$ such that:

- $E \subseteq U_e$ is the set of event identifiers. *Example:* the first event shown in Table 1 is related to the event identifier $e_1$.
- $AN \subseteq U_{att}$ is the set of attributes names. *Example:* in Table 1 *resource*, *prepaid-amount*, *weight*, and *total-weight* are attribute names and, in Table 2, *costs*, *color*, and *size* are attribute names.
- $AV \subseteq U_{val}$ is the set of attribute values (with the requirement that AN ∩ AV = ∅). *Example:* in Table 1 *200.0*, *Anahita*, and *10.0* are attribute values, and in Table 2, *Apple*, *green*, and *3500.0* are attribute values.
- $AT \subseteq U_{typ}$ is the set of attribute types. *Example:* the type of the attribute *resource* in Table 1 is *string*.
- $OT \subseteq U_{ot}$ is the set of object types. *Example:* in Table 2, for the first object, the type is *order*.
- $O \subseteq U_o$ is the set of object identifiers. *Example:* the first object in Table 2 is related to the object identifier $o_1$.
- $\pi_{typ} : AN \cup AV \rightarrow AT$ is the function associating an attribute name or value to its corresponding type. *Example:* for the attributes in Table 1, $\pi_{typ}(prepaid\text{-}amount) = float$, $\pi_{typ}(200.0) = float$.
- $\pi_{act} : E \rightarrow U_{act}$ is the function associating an event (identifier) to its activity. *Example:* for the first event shown in Table 1, the activity is *place order*.
- $\pi_{time} : E \rightarrow U_{timest}$ is the function associating an event (identifier) to a timestamp. *Example:* for the first event shown in Table 1, the timestamp is *2020-07-09T08:21:01.527+01:00*.
- $\pi_{vmap} : E \rightarrow (AN \nrightarrow AV)$ such that

  $\pi_{typ}(n) = \pi_{typ}(\pi_{vmap}(e)(n))$
  $\forall e \in E \; \forall n \in \mathrm{dom}(\pi_{vmap}(e))$

  is the function associating an event (identifier) to its attribute value assignments. *Example:* for the first event in Table 1, $\pi_{vmap}(e_1)(prepaid\text{-}amount) = 200.0$
- $\pi_{omap} : E \rightarrow \mathcal{P}(O)$ is the function associating an event (identifier) to a set of related object identifiers. *Example:* the first event in Table 1 is related to three objects $\pi_{omap}(e_1) = \{o_1, i_1, i_2\}$.

- $\pi_{otyp} \in O \rightarrow OT$ assigns precisely one object type to each object identifier. *Example:* for the first object in Table 2, $\pi_{otyp}(o_1) = order$.
- $\pi_{ovmap} : O \rightarrow (AN \nrightarrow AV)$ such that

  $\pi_{typ}(n) = \pi_{typ}(\pi_{ovmap}(o)(n))$
  $\forall n \in \mathrm{dom}(\pi_{ovmap}(o)) \; \forall o \in O$

  is the function associating an object to its attribute value assignments. *Example:* for the second object in Table 2, $\pi_{ovmap}(i_2)(color) = green$.
- $\leq$ is a total order (i.e., it respects the antisymmetry, transitivity, and connexity properties).

Recently, the OCEL format has been proposed for object-centric event logs[3]. Two implementations of the format exist (JSON-OCEL, supported by JSON; XML-OCEL, supported by XML; MongoDB [5]), with tool support available for some popular languages (Java/ProM framework[4], Javascript[5], Python[6]). On the page *Event Logs*, some event logs (in the JSON-OCEL and XML-OCEL formats) are available, which can be ingested by the tool support.

In Def. 8, some general statistics on object-centric event logs are introduced. While the number of events and (unique) objects derives directly from the log elements, the number of total objects is an interesting aggregation that considers how many events are related to the given object. So, considering the ratio between the number of total objects and unique objects, the higher the ratio, the higher the average length of the lifecycle of the objects of the object-centric event log.

**Definition 8** (General Statistics on an Object-Centric Event Log) Let $L$ be an object-centric event log as in Def. 7. We define the following general statistics on the object-centric event log:

GS1 *Number of Events* $\mathbb{E}(L) = |E|$.
GS2 *Number of Unique Objects* $\mathbb{UO}(L) = |O|$.
GS3 *Number of Total Objects* $\mathbb{TO}(L) = \sum_{e \in E} |\pi_{omap}(e)|$.

An operation defined on object-centric event logs is *flattening*. A flattening operation transforms the object-centric event log into a traditional event log given the choice of an object type. This is useful because many process mining approaches are only available for traditional event logs. Moreover, some object-centric process discovery algorithms (such as MVP [6,7] and object-centric Petri nets [8]) performs flattening to apply classic process discovery techniques and then unite the results for the different object types in a single model. Def. 9 proposes a formal definition of flattening. This is based on the definition of *restriction* for a function. Given a function $f : X \rightarrow Y$, and $X' \subseteq X$, $f|_{X'}$ is a function with $\text{dom}(f|_{X'}) = X'$ and for all $x \in X'$, $f(x) = f|_{X'}(x)$.

---

[3] http://www.ocel-standard.org/
[4] https://svn.win.tue.nl/repos/prom/Packages/OCELStandard/Trunk/
[5] https://github.com/Javert899/pm4js-sandbox
[6] https://github.com/OCEL-standard/ocel-support

**Definition 9** (Flattening with an Object Type) Let $L$ be an object-centric event log as in Def. 7, and $ot \in OT$ be an object type. We define the flattening of $L$ using ot as $FL(L, ot) = (E^{ot}, \pi_{act}^{ot}, \pi_{time}^{ot}, \pi_{case}^{ot}, \leq^{ot})$ where:

- $E^{ot} = \{e \in E \mid \exists_{o \in O} \pi_{otyp}(o) = ot \wedge o \in \pi_{omap}(e)\}$
- $\pi_{act}^{ot} = \pi_{act}|_{E^{ot}}$
- $\pi_{time}^{ot} = \pi_{time}|_{E^{ot}}$
- For $e \in E^{ot}$, $\pi_{case}^{ot}(e) = \{o \in \pi_{omap}(e) \mid \pi_{otyp}(o) = ot\}$
- $\leq^{ot} = \{(e_1, e_2) \in \leq \mid \exists_{o \in O} \pi_{otyp}(o) = ot \wedge o \in \pi_{omap}(e_1) \cap \pi_{omap}(e_2)\}$

Given the definition of flattening, we can introduce the notion of *lifecycle* in Def. 10. The lifecycle of an object is the corresponding case in the flattened log[7].

**Definition 10** (Lifecycle, Start and End Event for an Object) Let $L$ be an object-centric event log as in Def. 7. We define:

- The *lifecycle* of an object $o \in O$ as the sequence of events to which the object is related: $\text{lif}(o) = \text{case}_{FL(L, \pi_{otyp}(o))}(o)$
- The *trace* of an object $o \in O$ as the sequence of activities of the events belonging to its lifecycle: $\text{trace}(o) = \text{trace}_{FL(L, \pi_{otyp}(o))}(o)$
- The start activity of an object $o \in O$ as the first activity of its trace: $\text{start}(o) = \text{start}_{FL(L, \pi_{otyp}(o))}(o)$
- The end activity of an object $o \in O$ as the last activity of its trace: $\text{end}(o) = \text{end}_{FL(L, \pi_{otyp}(o))}(o)$

In Def. 10, we introduce the additional concepts of *trace* for an object (the activities of the events of its lifecycle). The start and end activities are of particular importance, as they are the start/end of the process execution and can be used to identify incomplete/improperly terminated objects.

## 3.2 Filtering

Filtering is an operation of high importance because it restricts the behavior contained in the log to the desired one. Many filters have been defined for traditional event logs (filtering the cases containing an activity, filters the cases starting or ending with an activity, timeframe filter). In this section, we want to introduce some filtering operations on object-centric event logs. In Def. 11, given a subset of events of the log, we define filtering operations restricting the event log to these events.

**Definition 11** (Filtering on a Set of Events) Let $L$ be an object-centric event log as in Def. 7, and $E' \subseteq E$ a set of events. We define the filtered event log $L_{E=E'} = (E', AN, AV, AT, OT, O, \pi_{typ}, \pi_{act}|_{E'}, \pi_{time}|_{E'}, \pi_{vmap}|_{E'}, \pi_{omap}|_{E'}, \pi_{otyp}, \pi_{ovmap}, \leq)$

Some filters based on Def. 11 are presented in Def. 12. These include a filter on a subset of activities (useful to remove some undesired activities from the analysis) and a filter on timeframe (useful to restrict the analysis to a given period of time).

**Definition 12** (Filtering on a Set of Events - Approaches) Let $L$ be an object-centric event log as in Def. 7. Let $A = \{\pi_{act}(e) \mid e \in E\}$ be the set of activities of $L$. We present some possibilities for the filtering of a set of objects:

F1 *Filtering on a Subset of Activities* Given a set of activities $A' \subseteq A$, filter on the events having an activity in $A'$: $E' = \{e \in E \mid \pi_{act}(e) \in A'\}$
F2 *Filtering on Timeframe* Given some lower and upper bounds lb, ub $\in U_{timest}$, filter on the events falling in the range [lb, ub]: $E' = \{e \in E \mid \text{lb} \leq \pi_{time}(e) \leq \text{ub}\}$

The filtered log is defined starting from $E'$ as in Def. 11.

It is also possible to define a filtering operation starting from a subset of objects. In Def. 13, the event log is filtered to the set of events that are related to one of these objects.

**Definition 13** (Filtering on a Set of Objects) Let $L$ be an object-centric event log as in Def. 7, and $O' \subseteq O$ a set of objects. Let $E_{O'} = \{e \in E \mid \pi_{omap}(e) \cap O' \neq \emptyset\}$ be the subset of events in $E$ related to at least one object in $O'$. We define the filtered event log
$L_{E=E_{O'}, O=O'} = (E_{O'}, AN, AV, AT, OT, O', \pi_{typ}, \pi_{act}|_{E_{O'}}, \pi_{time}|_{E_{O'}}, \pi_{vmap}|_{E_{O'}}, \pi_{omap}|_{E_{O'}}, \pi_{otyp}|_{O'}, \pi_{ovmap}|_{O'}, \leq)$

Some filters based on Def. 13 are presented in Def. 14. These exploit the operations on object-centric event logs introduced in Def. 10.

**Definition 14** (Filtering on a Set of Objects - Approaches) Let $L$ be an object-centric event log as in Def. 7. Let $A = \{\pi_{act}(e) \mid e \in E\}$ be the set of activities of $L$. We present some possibilities for the filtering of a set of objects:

F3 *Filtering on the Objects related to an Activity* Given a set of activities $A' \subseteq A$, filter on the objects related to one of these activities: $O' = \{o \in O \mid \text{trace}(o) \cap A' \neq \emptyset\}$

---
[7] So, is the sequence of events that are related to the object.

F4 *Filtering on Start Activities* Given a set of activities $A' \subseteq A$, filter on the objects starting with one of these activities: $O' = \{o \in O \mid \text{start}(o) \in A'\}$

F5 *Filtering on End Activities* Given a set of activities $A' \subseteq A$, filter on the objects ending with one of these activities: $O' = \{o \in O \mid \text{end}(o) \in A'\}$

F6 *Filter on a Path* Given a couple of activities $(a_1, a_2) \in A \times A$, filter on the objects containing $(a_1, a_2)$ in their trace: $O' = \{o \in O \mid (a_1, a_2) \in \text{trace}(o)\}$

F7 *Filter on Object Types* Given a set of object types $\text{SOT} \subseteq OT$, filter on the objects having one of these types: $O' = \{o \in O \mid \pi_{otyp}(o) \in \text{SOT}\}$

The filtered log is defined starting from $O'$ as in Def. 13.

With the approaches presented in Def. 12 and Def. 14, many filters available in the classic setting (endpoints, timeframe, attributes) are also made available in the object-centric setting.

### 3.3 Process discovery - OC-DFG

Some of the proposed approaches for object-centric process discovery on object-centric event logs follow a common schema: the object-centric event log is flattened on the available object types, a process model is discovered for the flattened logs, and then the results are collated together.

In this section, we formalize one object-centric process model, the *object-centric directly-follows multigraph* (OC-DFG), and how to discover an object-centric directly-follows multigraph starting from an object-centric event log. We choose to present object-centric directly-follows multigraphs in the context of the current section because:

- They can be straightforwardly discovered from object-centric event logs (flattening - discovery of DFG - collating).
- They can be easily annotated, given that no replay operation is necessary.

The formal definition of OC-DFG is presented in Def. 15. We see that an OC-DFG is a collection of nodes (the activities, plus one start and end node for each object type) and typed edges between the activities.

**Definition 15** (Object-Centric Directly-Follows Multigraph) An object-centric directly-follows multigraph (OC-DFG) is a tuple $(A, OT, N, F, \pi_{freqn}, \pi_{freqe})$ where:

- $A$ is a set of activities.
- $OT$ is a set of object types.
- $N = A \cup \{n_{S,\text{ot}} \mid ot \in OT\} \cup \{n_{E,\text{ot}} \mid ot \in OT\}$ is the set of nodes of the graph, which includes the set of activities and a start/end node for each object type ($n_{S,\text{ot}}$ and $n_{E,\text{ot}}$ respectively).
- $F \subseteq N \times OT \times N$ is a set of typed arcs.
- $\pi_{freqn} : A \nrightarrow \mathbb{N}$ assigns a frequency to the activities.
- $\pi_{freqe} : F \nrightarrow \mathbb{R}^+ \cup \{0\}$ assigns a frequency to the arcs.

Def. 16 introduces the discovery of OC-DFG from object-centric event logs. Essentially, the event log is flattened for each object type, the operation of discovery of an object-centric directly-follows multigraph is performed for each flattened log and the results are collated together to obtain the OC-DFG.

**Definition 16** (Discovery of an Object-Centric Directly-Follows Multigraph) Let $L$ be an object-centric event log as in Def. 7. We define $ODFG(L) = (A, OT, N, F, \pi_{freqn}, \pi_{freqe})$ where $A$ and $OT$ are the set of activities and object types of the log, respectively, $N$ is obtained as in Def. 15, and given $n_1, n_2 \in N$ we have $(n_1, ot, n_2) \in F \iff (n_1, n_2) \in \pi_F(DFG(FL(L, ot)))$, and $\text{dom}(\pi_{freqn}) = \text{dom}(\pi_{freqe}) = \emptyset$ (no frequency is described in this definition).

Figure 1 shows an example object-centric directly-follows multigraph. The example contains different object types and tells some information about the lifecycles of the different object types, including:

- The lifecycle of the objects with type *DOCTYPE_Inquiry* starts and ends with the activity "Create Quotation".
- The lifecycle of the objects with type *DOCTYPE_Quotation* starts with a "Create Quotation" activity, which can end the lifecycle of the quotation or lead to the "Create Order" activity.
- The lifecycle of the objects with type *DOCTYPE_Order* allows for a "Create Order" activity, which can end the lifecycle of the order or lead to the "Create Goods Movement" activity.

Def. 17 defines a frequency-based filtering on object-centric event logs. This is useful to simplify the model after the discovery, for example, by focusing on the mainstream behavior (most frequent activities and edges).

**Definition 17** (Frequency-Based Filtering) Let $(A, OT, N, F, \pi_{freqn}, \pi_{freqe})$ be an object-centric directly-follows multigraph. Given $m_n$ and $m_e$, which are thresholds for the frequencies of the activities and edges respectively, we define the filtered object-centric directly-follows multigraph $(A', OT, N', F', \pi'_{freqn}, \pi'_{freqe})$, where $A' = \{a \in A \mid \pi_{freqn}(a) \geq m_n\}$, $N' = A' \cup \{n_{S,\text{ot}} \mid ot \in OT\} \cup \{n_{E,\text{ot}} \mid ot \in OT\}$, $F' = \{f \in F \mid \pi_{freqe}(f) \geq m_e\} \cap (N' \times OT \times N')$, $\pi'_{freqn} = \pi_{freqn}|_{A'}$, $\pi'_{freqe} = \pi_{freqe}|_{F'}$.

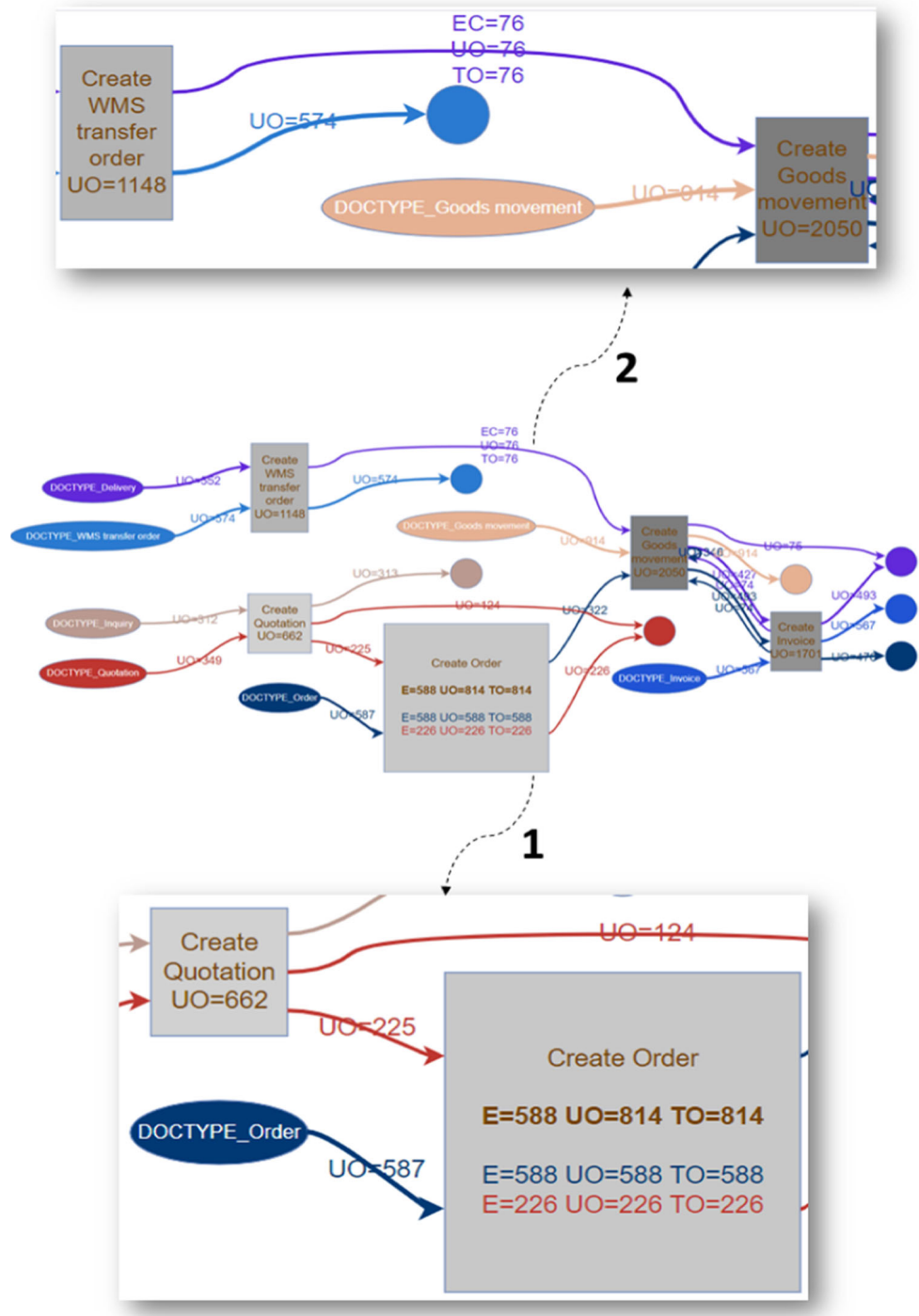

**Fig. 1** An example object-centric directly-follows multigraph (complete view). The arc with number 1 highlights the activity "Create Order", which shows all the statistics for all the object types of the event log. The arc with number 2 highlights the arc going from "Create WMS transfer order" to "Create Goods movement", which shows the statistics for the three annotations (E/O/EC)

We see that in Def. 16 we do not introduce any frequency measure on the nodes/edges of the OC-DFG. In Def. 18 and Def. 19, some frequency metrics are introduced for activities and paths, respectively, and the discovery of OC-DFGs can be modified with the inclusion of these measures. It should be noted that the type of models presented in [7] is equivalent to OC-DFGs with the choice of AF1 as frequency metric for the activities and PF2 as frequency metric for the paths.

### 3.4 Activity/Path metrics

This section proposes some frequency metrics for activities and paths that can be computed starting from an object-centric event log. The measures are independent from the type of model but can be used to annotate the model (for example, the OC-DFGs introduced in Def. 15).

Def. 18 proposes some metrics at the activity level (number of events, number of unique objects, number of total objects).

**Definition 18** (Activity Frequency Metrics) Let $L$ be an object-centric event log as in Def. 7. Let $A = \{\pi_{act}(e) \mid e \in E\}$ be the set of activities of $L$, and for $a \in A$, the set $E_a = \{e \in E \mid \pi_{act}(e) = a\}$ of all the events having $a$ as activity. We define the following metrics on an activity $a \in A$:

AF1 *Counting the Number of Events having a given Activity* $\mathbb{E}(a) = |E_a|$.
AF2 *Counting the Number of Unique Objects related to Events having a given Activity* $\mathbb{UO}(a) = |\{o \in O \mid \exists e \in E_a, o \in \pi_{omap}(e)\}|$
AF3 *Counting the Number of Total Objects related to Events having a given Activity* $\mathbb{TO}(a) = |\{(e, o) \in O \mid e \in E_a \wedge o \in \pi_{omap}(e)\}|$

The metrics proposed in Def. 18 can be applied either on the overall log, or on the log filtered on a specific object type (see filter 7 of Def. 14). Figure 1(1) shows the annotations (extracted using the tool proposed in Sect. 4) on the activity "Create Order". In the same box, we have different lines:

- The name of the activity.
- The three frequency annotations ($\mathbb{E}$, $\mathbb{UO}$, $\mathbb{TO}$) on the overall log.
- The three frequency annotations on the log filtered on the *DOCTYPE_Order* object type (which is colored blue).
- The three frequency annotations on the log filtered on the *DOCTYPE_Quotation* object type (which is colored red).

Def. 19 defines some metrics on the paths based on the lifecycle of the objects.

**Definition 19** (Paths Frequency Metrics) Let $L$ be an object-centric event log as in Def. 7. Let $A = \{\pi_{act}(e) \mid e \in E\}$ be the set of activities of $L$, and for $a \in A$, the set $E_a = \{e \in E \mid \pi_{act}(e) = a\}$ of all the events having $a$ as activity. Let $ot \in OT$ be an object type, and $O_{ot} = \{o \in O \mid \pi_{otyp}(o) = ot\}$ be the set of all the objects having object type ot. We define the following metrics, depending on ot, provided two activities $a_1, a_2 \in A$:

PF1 *Counting the Number of Event Couples which realize the Path* $\mathbb{EC}(a_1, ot, a_2) = |\{(e_1, e_2) \in E_{a_1} \times E_{a_2} \mid \exists o \in O_{ot}, (e_1, e_2) \in \text{lif}(o)\}|$.
PF2 *Counting the Number of Objects having the Path in their Lifecycle* $\mathbb{UO}(a_1, ot, a_2) = |\{o \in O_{ot} \mid (a_1, a_2) \in \text{trace}(o)\}|$.
PF3 *Counting the Number of Total Objects having the Path in their Lifecycle* $\mathbb{TO}(a_1, ot, a_2) = |\{(e_1, o, e_2) \in E_{a_1} \times O_{ot} \times E_{a_2} \mid (e_1, e_2) \in \text{lif}(o)\}|$.

Figure 1(2) shows the annotations (extracted using the tool proposed in Sect. 4) on the path of type *DOCTYPE_Delivery* between the activities "Create WMS transfer order" and "Create Goods movement". We see that the three proposed measures ($\mathbb{EC}$, $\mathbb{UO}$, $\mathbb{TO}$) are all reported.

As the last technique, we describe some approaches for conformance checking which are independent of a process model and depend solely on the verification of properties on the event log. Def. 20 presents formally the rules.

**Definition 20** (Conformance Checking - Model Independent Approaches) Let $L$ be an object-centric event log as in Def. 7. Let $A = \{\pi_{act}(e) \mid e \in E\}$ be the set of activities of $L$, and $E_a = \{e \in E, \pi_{act}(e) = a\}$. We present some possibilities for conformance checking on top of object-centric event logs:

CC1 *Number of Objects related to an Activity* for $a \in A$, we define for lb, ub $\in \mathbb{N}$ (lower and upper bound for the number of related objects)

$$\text{conf}_{\text{num\_obj}}(a, \text{lb}, \text{ub}) = \{e \in E_a \mid |\pi_{omap}(e)| < \text{lb} \vee |\pi_{omap}(e)| > \text{ub}\}$$

as the set of events violating the rule.

CC2 *Duration of the Lifecycle* given lb, ub $\in \mathbb{R}^+ \cup \{0\}$, we define:

$$\text{conf}_{\text{dur\_lif}}(\text{lb}, \text{ub}) = \{o \in O \mid \\ \pi_{time}(\text{lif}(o)(|\text{lif}(o)|)) - \pi_{time}(\text{lif}(o)(1)) < \text{lb} \vee \\ \pi_{time}(\text{lif}(o)(|\text{lif}(o)|)) - \pi_{time}(\text{lif}(o)(1)) > \text{ub}\}$$

as the set of objects violating the rule.

The rule CC1 is useful to identify situations where an excessive number of objects is worked by an activity. For example, we can think of an activity "Approve Expense Report" which usually involves a limited number of expense reports. If an event with the activity "Approve Expense Report" involves 50 different reports, CC1 is useful to identify the given event. The rule CC2 is helpful in identifying objects with an extremely long lifecycle. As an example, if a ticket is supposed to be approved in one week, while it is still not closed after one month, CC2 is useful for identifying the given ticket (object).

Figure 2 proposes a visualization of the rule CC1 and Fig. 3 provides a visualization of the rule CC2. For both, we assume that the event log is filtered on the different object types, and the rules are applied to the filtered logs. For CC1, we calculate for each activity the average $\mu$ and the standard deviation $\sigma$ of the number of related objects for the events of the given activity. Then, we assume that every event of the given activity having a number of related objects that is lower than $\mu - \zeta * \sigma$ or higher than $\mu + \zeta * \sigma$ (where $\zeta \in \mathbb{R}^+ \cup \{0\}$ is a positive number) are anomalous[8]. For CC2, we calculate for each object the average $\mu$ and the standard deviation $\sigma$ of the duration of the lifecycle of the object. Then, we assume that every object having a duration of the lifecycle that is lower than $\mu - \zeta * \sigma$ or higher than $\mu + \zeta * \sigma$ (where $\zeta \in \mathbb{R}^+ \cup \{0\}$ is a positive number) are anomalous[9]. For both rules, it is possible to filter the object-centric event log, keeping respectively only the anomalous events and the anomalous objects.

| Activity | Object Type | Avg Objects | Std Objects | Num. D |
|---|---|---|---|---|
| place order | items | 5.103198921436451 | 2.31762423494253 | 1789 |
| place order | products | 12.45 | 0.8046738469715539 | 4 |
| confirm order | items | 5.103198921436451 | 2.31762423494253 | 1789 |
| confirm order | products | 12.45 | 0.8046738469715539 | 4 |
| pay order | products | 12.45 | 0.8046738469715539 | 4 |
| pay order | items | 5.103198921436451 | 2.31762423494253 | 1789 |
| create package | items | 8.096948155411201 | 3.80100544057014 | 2774 |
| create package | products | 13.95 | 0.21794494717703364 | 1 |
| create package | orders | 4.6925 | 1.567783068539776 | 741 |

**Fig. 2** Conformance checking based on the number of related objects (in the web-based tool proposed in Sect. 4.1)

| Object Type | Avg Lifecycle Dur.(s) | Stdev Lifecycle Dur.(s) | Number of De |
|---|---|---|---|
| customers | 36504286.52941176 | 1917740.6139664985 | 6 |
| items | 1313686.2029660498 | 1097726.4767986913 | 1124 |
| orders | 1678116.412 | 1171904.9000183828 | 371 |
| packages | 156197.43547169812 | 127195.17230464106 | 406 |
| products | 36810777.25 | 1931374.304741856 | 7 |

**Fig. 3** Conformance checking based on the objects lifecycle duration (in the web-based tool proposed in Sect. 4.1)

## 4 Tool

This section presents two tool supports for the process discovery techniques proposed in this paper: a web-based tool and an implementation on top of the ProM process mining framework. These are available and described at https://ocpm.info. Along with process discovery, both tools support flattening and filtering. In particular, the web-based tool also supports conformance checking and statistics on object-centric event logs.

---

[8] Choosing $\zeta = 1$ includes all the events for which the number of related objects is more deviant than one standard deviation from the average. Choosing $\zeta = 6$ includes all the events for which the number of related objects is more deviant than six standard deviations from the average.

[9] Choosing $\zeta = 1$ includes all the objects for which the duration of the lifecycle is more deviant than one standard deviation from the

Footnote 9 continued
average. Choosing $\zeta = 6$ includes all the objects for which the duration of the lifecycle is more deviant than six standard deviations from the average (six sigma principle).

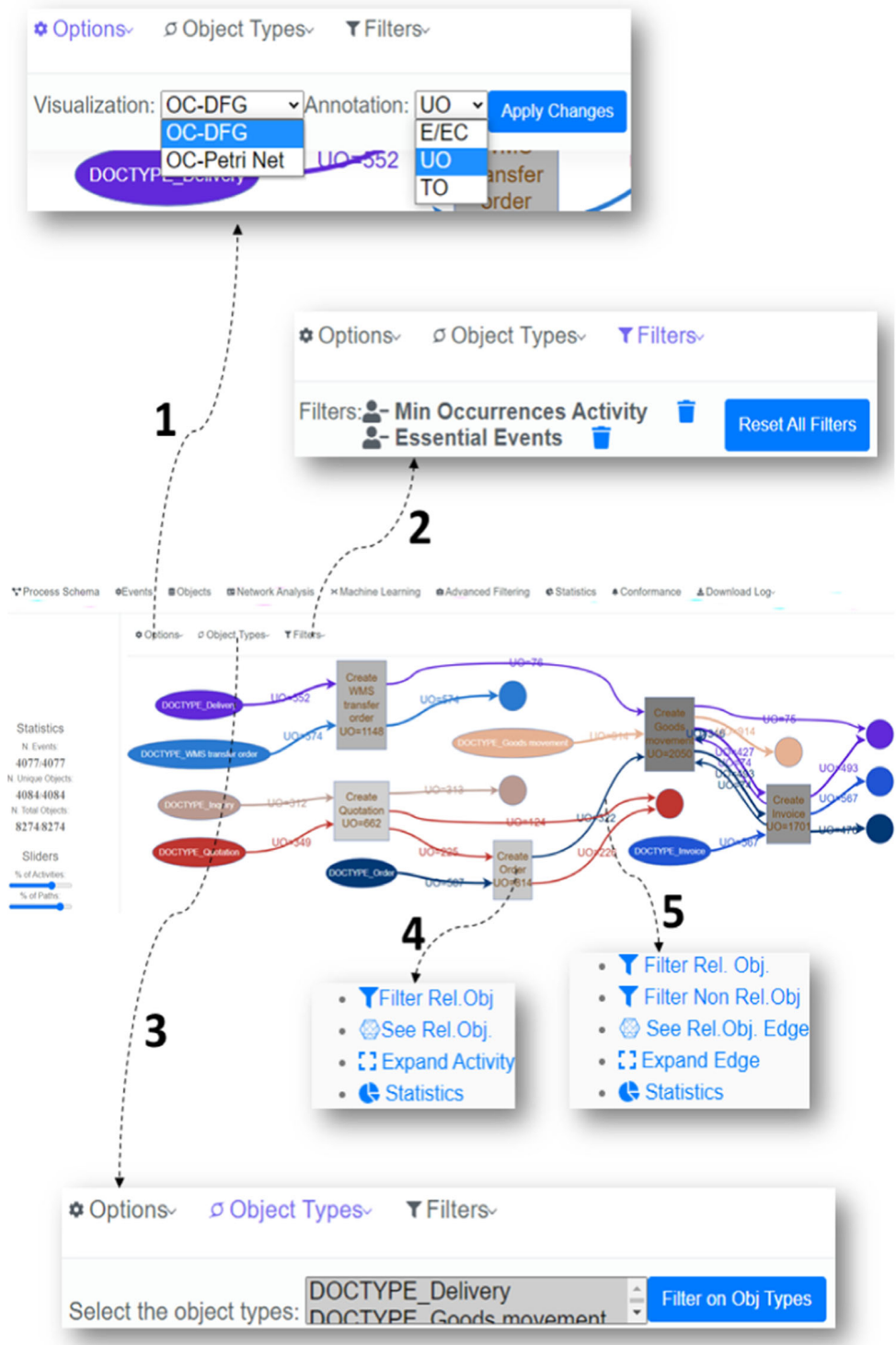

**Fig. 4** Overall view over the process model page of the proposed web-based tool

## 4.1 OC-PM (Web-based tool)

We present a novel tool for object-centric process mining, OC-PM, which enables the object-centric process mining analyses presented in this paper. The tool is available at the address https://ocpm.info and consists of HTML/Javascript content that can be downloaded and promptly run in the browser without any backend. The first page of the tool enables the upload of an object-centric event log in the JSON-OCEL or XML-OCEL formats[10].

---

[10] Some of these logs are available at the address http://www.ocel-standard.org/

The tool consists of different pages, including the *Process Schema* (which visualizes a process model with the possibility to interact with it), the *Events* page (which visualizes the list of events contained in the object-centric event log, with the possibility to focus on the events belonging to the lifecycle of an object), the *Objects* page (the list of objects of the log having a given object type is shown, along with their lifecycle, the duration of the lifecycle, and other statistics), the *Statistics* page (showing some generic graphs including the number of related objects per object type, the number of events per activity, the number of related objects per event, the distribution of the length of the lifecycle of objects, the distribution of the events during the time, the dotted chart), the *Conformance* page (providing some conformance checking functionalities).

Figure 4 shows the process model page of the proposed tool. The page is organized as follows:

- The top menu presents the different pages/features of the application.
- The second menu presents some options, shown in Fig. 4(1), including the type of the process model:
  - Object-centric directly-follows multigraphs (described in Sect. 3.3).
  - Object-centric Petri nets [8]. In particular, the decorations are obtained using the token-based replay technique described in [9].
  
  and the type of annotation:
  - With the *E/EC* option, the process model is annotated using the measure $\mathbb{E}$ for the activities and the measure $\mathbb{EC}$ for the edges.
  - With the *UO* option, the process model is annotated using the measure $\mathbb{UO}$ for the activities and the edges.
  - With the *TO* option, the process model is annotated using the measure $\mathbb{TO}$ for the activities and the edges.
  
  Moreover, the filtering on object types, along with a filters chain functionality (which shows the active filters, with the possibility to remove them), is implemented in this menu (see Fig. 4(2-3)).
- The left panel shows the number of events, unique objects, and total objects of the overall log (see Def. 8). Moreover, a *sliding* functionality is offered, keeping only the most frequent activities/edges (this is done for OC-DFGs using the approach described in Def. 17).
- The right panel shows the process model.

The process model page permits interaction with the activities and the edges. The filtering approaches F1-7 presented in Def. 14 are all implemented in the tool. Figure 4(4-5) shows the interaction menus when an activity and an edge are clicked, respectively. It is possible to apply the filtering, or to observe the list of objects related to the activity and edge.

Figure 5 and 6 show some of the statistics that can be computed on object-centric event logs. Among the additional features, the download of the filtered event log in the JSON-OCEL or XML-OCEL is available, and the possibility to flatten the object-centric event log to a traditional event log saved in the XES format is offered (Fig. 7).

### 4.2 OC-PM (ProM framework)

We present another implementation of the process discovery techniques proposed in this paper, on top of the popular process mining framework ProM 6.x[11]. The implementation is proposed in the package OCELStandard[12], which can be downloaded using the package manager of ProM. An object-centric event log, in the JSON-OCEL or XML-OCEL formats, can be imported in ProM using the import button on the top right. After importing, some object-centric process mining features are available on top of the object-centric event log: flattening to an object type and process discovery (object-centric directly-follows multigraphs and object-centric Petri nets[13]). Opening the process discovery plugin, a visualization of an object-centric directly-follows multigraph with a default choice for the activity/path sliders is proposed. The notation is analogous to the one of the web-based tool presented in the previous subsection. The user can interact with the diagram by clicking on the nodes (activities) and edges of the directly-follows multigraph. The values for the activity/path sliders can be changed on the top panel. The user can also apply some filters on the object-centric event log starting from the top panel.

## 5 Related work

This section presents the related work on object-centric process mining.

### 5.1 Artifact-centric approaches

Artifact-centric process mining is based on defining the properties of key business-relevant entities called business artifacts. In particular, the proposed techniques focus on the modeling of the single artifacts and their interactions. In [10], two-phases conformance checking approach is proposed, in which the conformance is checked both in the single artifacts both in the interactions between them. In [11], an approach to discover the artifacts and their lifecycle from a relational database is proposed. This is done by identifying the artifacts and extracting event logs for each artifact. In [12], the discovery of artifact-centric models on top of the SAP ERP system is discussed. A limitation of these approaches is the lack of comprehensive tool support and the dependence on a relational database schema.

---

[11] https://www.promtools.org/doku.php?id=prom611

[12] https://svn.win.tue.nl/repos/prom/Packages/OCELStandard/Trunk/

[13] The decorations are obtained using the token-based replay technique described in [9].

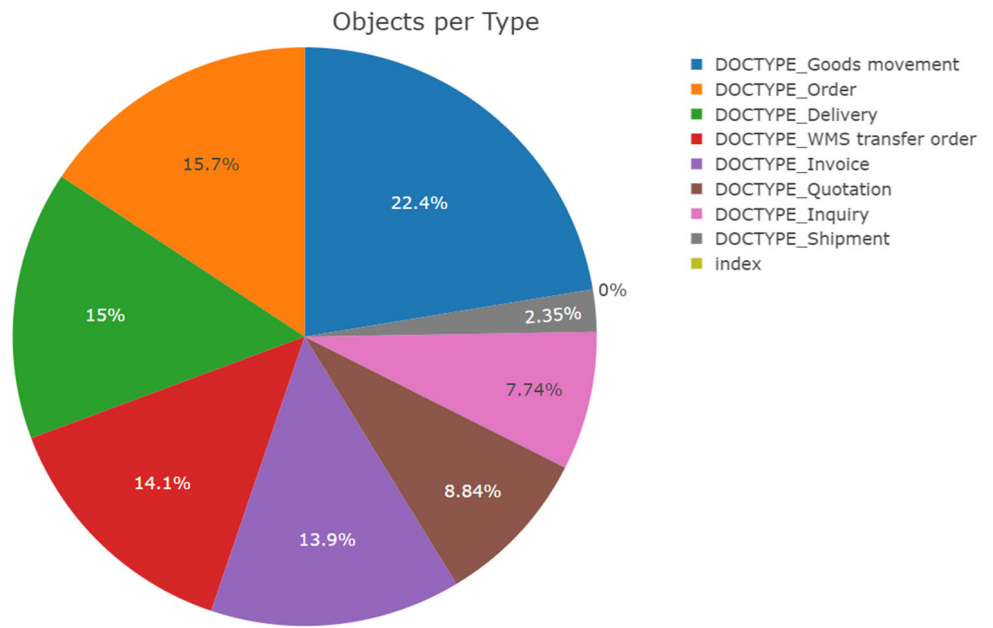

**Fig. 5** Statistic proposed in the web-based tool: number of objects per type

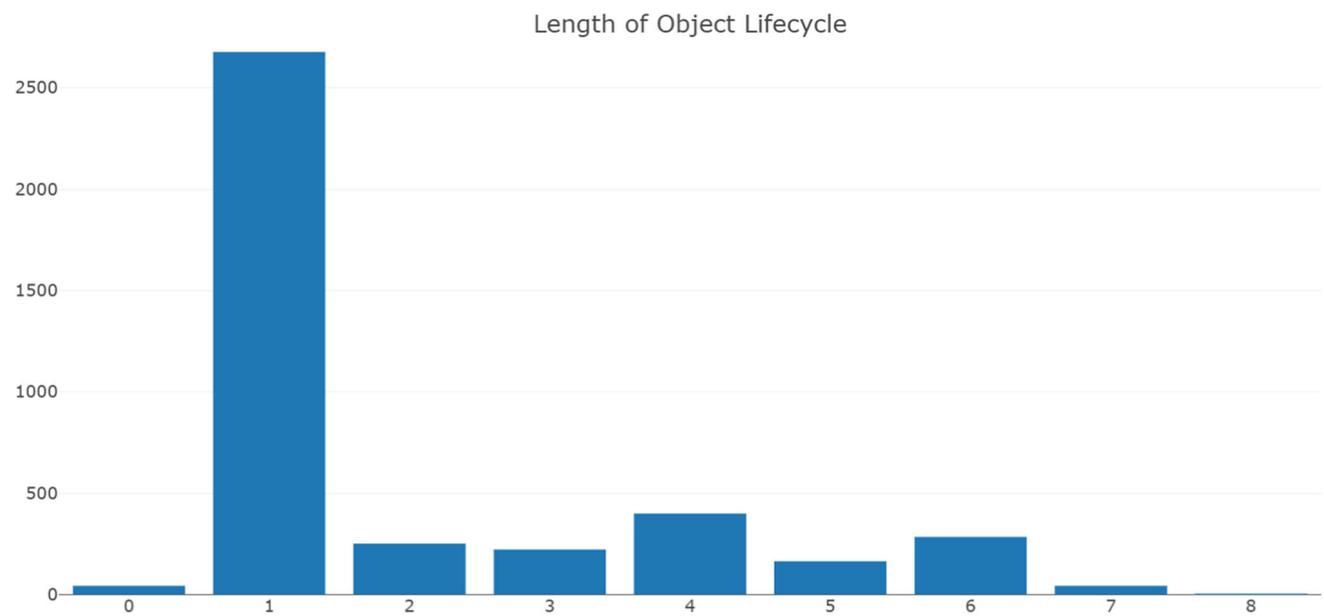

**Fig. 6** Statistic proposed in the web-based tool: length of the lifecycle

### 5.2 Object-centric behavioral constraint models

In [13], the *object-centric behavioral constraint models* (OCBC) are proposed as declarative models with rich semantics that can describe the interaction between the different entities of a database and the activities recorded in an object-centric event log with the features described in [14]. However, the discovery of the rich set of constraints and the proposed event log format (storing the entire state of the object model for each event) have scalability issues.

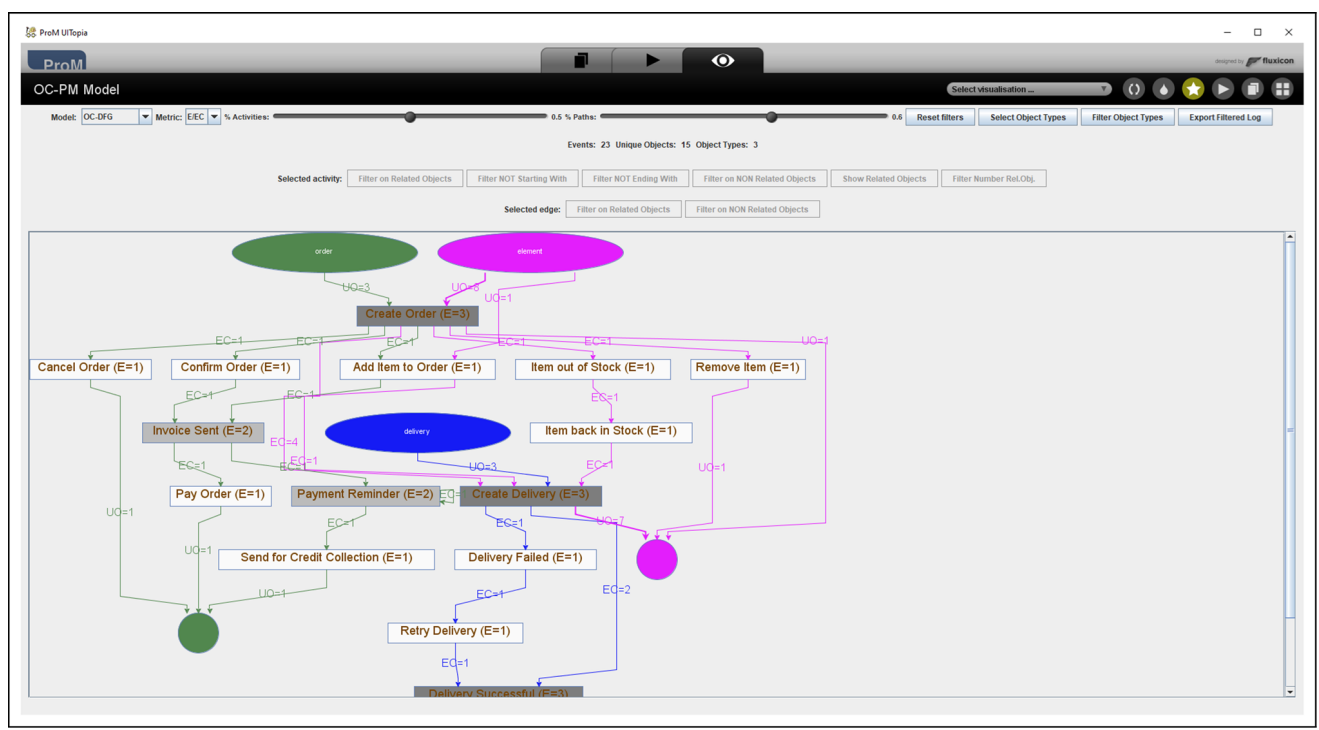

**Fig. 7** Implementation of the object-centric process discovery techniques as plug-in of the ProM process mining framework

## 5.3 Petri nets-based approaches

Colored Petri nets [15] have been proposed in the '80 and have a wide range of applications. Colored Petri nets allow the storage of a data value for each token. The data value is called the token color. Every place contains tokens of one type, which is referred to as the color set of the place. Moreover, expressions are defined at the arc level for consumption/production purposes, and some guards can control the execution of the transitions. Given their rich semantics, the proposal of a process discovery algorithm able to manage colors, color sets, expressions, and guards is an enormous challenge. In [16], colored Petri nets are extended (with the name catalog-nets) to accommodate processes with several cases that need to co-evolve flexibly.

In [17], three concepts are provided to describe the behavior of processes with many-to-many interactions:

- Unbounded dynamic synchronization of transitions.
- Cardinality constraints limit the size of the synchronization.
- Correlation of the token identities based on history.

## 5.4 Graph and process querying

In [18,19], the usage of graph databases for the storage, querying, and aggregation of object-centric event data is proposed. An object-centric event log is inserted in the graph by creating nodes for the events, objects, object types, attributes of the event log, and connections are created based on the content of the log. In [20], an algorithm for the discovery of directly-follows graphs on top of graph databases is proposed. However, the scalability of graph databases on process mining tasks still needs to be investigated thoroughly. In [20], the execution time of process mining tasks in a popular graph database (Neo4J) is shown to be disappointing.

In [21], a query language to analyze the execution of business processes is proposed. An approach for ontology-based extraction of event data has been proposed in [22].

## 5.5 Flattening-based process discovery

A discovery operation can be defined by flattening (see Def. 11) the object-centric event log into the different object types, discovering traditional process models (as an example, a DFG or a Petri net) on top of the flattened logs and then collating the results together. Different process models can serve as building blocks and have been proposed in the literature:

- *Object-centric directly-follows graphs*: in [7], the usage of object-centric directly-follows multigraphs is proposed to describe the activities of an object-centric event log, and the interactions between them.
- *Object-centric Petri nets*: in [8], object-centric Petri nets have been proposed to support a subset of the semantics

of colored Petri nets. A discovery approach is proposed starting from object-centric event logs, in which a flattened log is obtained for each object type, a mainstream process discovery algorithm (such as the alpha miner or the inductive miner) is applied on top of the flattened log, and the Petri nets are then collated together into an object-centric Petri net. In the model, every place and arc is associated with a unique object type, and an arc can be allowed to consume/produce a single or multiple tokens.

### 5.6 Other approaches

Some object-centric process models have been proposed in [23,24], however the tool support/assessment is lacking.

The relationship between interconnected processes has been investigated. In [25], a token-based interaction monitoring framework is proposed. In [26], instance-spanning constraints are discovered from event logs, which regulate the start of the process instances. In [27], object-state transitions are proposed to improve business process intelligence. Although all these approaches are useful for conformance checking, they do not result in a comprehensive process model. In [28], multi-instance mining has been proposed, along with an implementation in the ProM framework that can show the interaction between the states visually.

## 6 Conclusion

The current paper describes a set of object-centric process mining techniques which can be used to analyze object-centric event logs extracted from mainstream information systems (such as SAP ERP). The definition of object-centric event logs, and the introduction of the OCEL format, permit the introduction of some operations both at the formal level both in tools/libraries supporting OCEL. The operations of flattening (projecting the object-centric event log to a traditional event log after the choice of an object type) and filtering (activities, paths, endpoints, timeframe, object types) are important for the development of more advanced object-centric process mining techniques. In particular, the flattening operation is an essential operation for process discovery.

The paper also proposes an object-centric process model (the object-centric directly-follows multigraph, OC-DFG), which can be straightforwardly discovered from object-centric event logs, and easily annotated with frequency measures. Moreover, several frequency measures for the activities and the paths of the event log are introduced, which can be used as annotations for OC-DFGs and other types of object-centric process models. Eventually, some conformance checking approaches for object-centric event logs are introduced, which verify some properties of the events/objects of the log.

Comprehensive tool support, which is available as a web interface and as plugin in the ProM framework, is offered for the ingestion, exploration, and discovery of OC-DFGs and of object-centric Petri nets [8], statistical analysis, and conformance checking. To the best of the authors' knowledge, this is the first attempt to provide comprehensive tool support in the object-centric setting.

There are several points of interest in object-centric process mining not discussed in the current paper, including a more precise visualization of the interactions between different objects and model-based conformance checking. Moreover, an assessment of the proposed techniques on real-life event logs is missing from the current paper. As the discipline is still young, these points can be developed in future work.

**Funding** Open Access funding enabled and organized by Projekt DEAL.